\RequirePackage[2020-02-02]{latexrelease}
\documentclass[aps,prl,twocolumn,superscriptaddress]{revtex4}

\usepackage{amssymb}
\usepackage{amsmath}
\usepackage{wrapfig}
\usepackage{caption}
\usepackage{color}
\usepackage{xfrac}
\usepackage{wrapfig}

\newcommand{\degree}{$^\circ$}
\newcommand{\R}{\textcolor{black}}
\newcommand{\BB}{\textcolor{black}}
\newcommand{\ea}{\textit{et al.}}

\begin{document}



\title{Using Diffuse Scattering to Observe \\X-Ray-Driven Nonthermal Melting}

\author{N. J. Hartley}
\email{njh@slac.stanford.edu}
\affiliation{SLAC National Accelerator Laboratory, CA 94025, USA}
\affiliation{Helmholtz Zentrum Dresden Rossendorf, Dresden 01099, Germany}

\author{J. Grenzer}
\affiliation{Helmholtz Zentrum Dresden Rossendorf, Dresden 01099, Germany}

\author{L. Huang}
\affiliation{Helmholtz Zentrum Dresden Rossendorf, Dresden 01099, Germany}
\author{Y. Inubushi}
\affiliation{Japan Synchrotron Radiation Research Institute, Sayo, Hyogo 679-5198, Japan}
\affiliation{RIKEN Spring-8 Center, Sayo, Hyogo 679-5148, Japan}
\author{N. Kamimura}
\affiliation{Graduate School of Engineering, Osaka University, Suita, Osaka 565-0087, Japan}
\author{K. Katagiri}
\affiliation{Graduate School of Engineering, Osaka University, Suita, Osaka 565-0087, Japan}
\author{R. Kodama}
\affiliation{Graduate School of Engineering, Osaka University, Suita, Osaka 565-0087, Japan}
\affiliation{Photon Pioneers Center, Osaka University, Suita, Osaka 565-0087, Japan}
\author{A. Kon}
\affiliation{Japan Synchrotron Radiation Research Institute, Sayo, Hyogo 679-5198, Japan}
\affiliation{RIKEN Spring-8 Center, Sayo, Hyogo 679-5148, Japan}
\author{W. Lu}
\affiliation{European XFEL GmbH, Holzkoppel 4, D-22869 Schenefeld, Germany}
\author{M. Makita}
\affiliation{European XFEL GmbH, Holzkoppel 4, D-22869 Schenefeld, Germany}
\author{T. Matsuoka}
\affiliation{Graduate School of Engineering, Osaka University, Suita, Osaka 565-0087, Japan}
\author{S. Nakajima}
\affiliation{Graduate School of Engineering, Osaka University, Suita, Osaka 565-0087, Japan}
\author{N. Ozaki}
\affiliation{Graduate School of Engineering, Osaka University, Suita, Osaka 565-0087, Japan}
\affiliation{Photon Pioneers Center, Osaka University, Suita, Osaka 565-0087, Japan}
\author{T. Pikuz}
\affiliation{Graduate School of Engineering, Osaka University, Suita, Osaka 565-0087, Japan}
\author{A. Rode}
\affiliation{Laser Physics Centre, Research School of Physical Science and Engineering, Australian National University, Canberra, ACT0200, Australia}
\author{D. Sagae}
\affiliation{Graduate School of Engineering, Osaka University, Suita, Osaka 565-0087, Japan}
\author{A. K. Schuster}
\affiliation{Helmholtz Zentrum Dresden Rossendorf, Dresden 01099, Germany}
\affiliation{Technische Universit{\"a}t Dresden, 01062 Dresden, Germany}
\author{K. Tono}
\affiliation{Japan Synchrotron Radiation Research Institute, Sayo, Hyogo 679-5198, Japan}
\affiliation{RIKEN Spring-8 Center, Sayo, Hyogo 679-5148, Japan}
\author{K. Voigt}
\affiliation{Helmholtz Zentrum Dresden Rossendorf, Dresden 01099, Germany}
\affiliation{Technische Universit{\"a}t Dresden, 01062 Dresden, Germany}
\author{J. Vorberger}
\affiliation{Helmholtz Zentrum Dresden Rossendorf, Dresden 01099, Germany}
\author{T. Yabuuchi}
\affiliation{Japan Synchrotron Radiation Research Institute, Sayo, Hyogo 679-5198, Japan}
\affiliation{RIKEN Spring-8 Center, Sayo, Hyogo 679-5148, Japan}

\author{E. E. McBride}
\affiliation{SLAC National Accelerator Laboratory, CA 94025, USA}
\author{D. Kraus}
\affiliation{Helmholtz Zentrum Dresden Rossendorf, Dresden 01099, Germany}
\affiliation{Institut für Physik, Universität Rostock, Albert-Einstein-Str. 23, 18059 Rostock, Germany}

\date{\today}

\begin{abstract}
We present results from the SPring-8 Angstrom Compact free electron LAser (SACLA) XFEL facility, using a high intensity ($\sim\!10^{20}\,$W/cm$^2$) X-ray pump X-ray probe scheme to observe changes in the ionic structure of silicon induced by X-ray heating of the electrons. By avoiding Laue spots in the scattering signal from a single crystalline sample, we observe a rapid rise in diffuse scattering and a transition to a disordered, liquid-like state with a significantly different structure to liquid silicon. The disordering occurs within 100 fs of irradiation, a timescale which agrees well with first principles simulations, and is faster than that predicted by purely inertial behavior, suggesting that both the phase change and disordered state reached are dominated by Coulomb forces. This method is capable of observing liquid scattering without masking signal from the ambient solid, allowing the liquid structure to be measured throughout and beyond the phase change.
\end{abstract}

\maketitle



Nonthermal melting is the loss of periodic order in a system without thermal equilibration between the electron and ion subsystems. It has been observed in a wide range of semiconductors \cite{Sokolowski-Tinten1998,Rousse2001,Siders1999}, where a high-fluence laser pulse can excite large numbers of electrons to the conduction band. This leads to an electrostatic potential between the now-ionized atoms and, if the crystalline structure is open-packed, rapid disordering into a closer-packed state, a process known as Crystal Mismatch Heating (CMH) \cite{Lyon2015}. This process, or the closely-related Coulomb explosion \cite{Medvedev2013a}, is therefore expected in any system where a large population of electrons can be excited on ultrafast timescales. While optical or infrared photons can cause such excitation in semiconductors, high fluence X-ray pulses, capable of penetrating through samples and exciting core electrons, can cause such an effect in a much wider range of materials \cite{Medvedev2018a}.

X-ray Free Electron Lasers (XFELs) can deliver far higher brilliances than previous X-ray sources, and together with new focusing approaches that deliver spot sizes of less than 1 $\mu$m in diameter \cite{Yumoto2013,Yumoto2020}, irradiation intensities comparable to those of optical lasers can now be reached. Unlike in optical-driven transitions, X-rays can deposit energy throughout thick samples, exciting high-energy photoelectrons which take more time to thermalize with the electron system, heating electrons over a larger region. Although this means that a lower energy density will be reached with the same incident intensity, it delivers a significantly more homogeneous initial energy density across a larger volume than can be excited with optical lasers.

To date, X-ray measurements of ultrafast melting or disordering have largely relied on changes in diffraction peaks, as in previous work at SACLA \cite{Hartley2019, Inoue2016} and elsewhere \cite{Pardini2018}. The latter, by Pardini \ea, demonstrates the challenges of studying phase transitions by observing weakening of diffraction lines, particularly when both the pump and probe pulse are at the same energy; what was initially believed to be strong diffraction signal from \BB{the silicon 333 reflection} up to 150 fs after X-ray heating was in fact due to fluctuations in the pulse fluences and a loss of beam overlap \cite{Pardini2018, Pardini2020}. The work we present here also looks at X-ray induced silicon melting, but with our single crystal sample aligned such that none of the Laue spots meet the diffraction condition. The transition from solid to liquid can then be discerned from the rise in diffusely scattered signal as the lattice order is lost, as has previously been seen in optically driven semiconductors \cite{Lindenberg2008} and structural transitions \cite{Wall2018}, rather than from the drop in the diffraction signal.

\vspace{5mm}
\begin{figure}
	\includegraphics[width=0.48\textwidth]{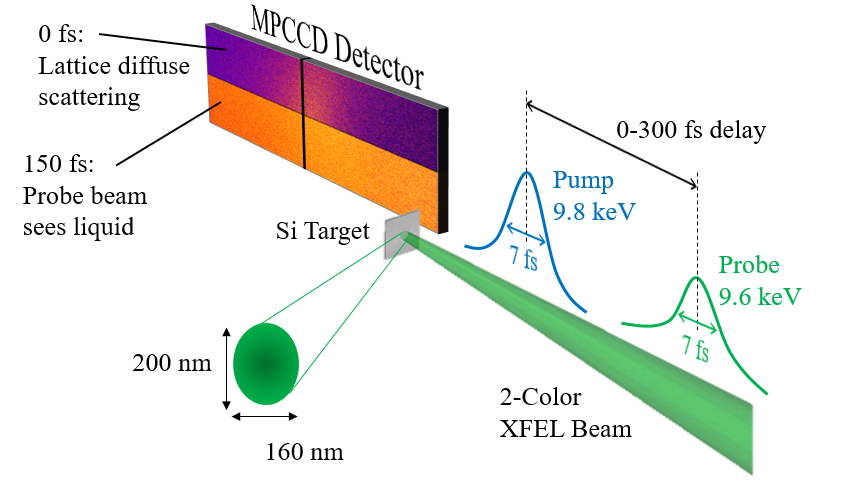} 
	\captionsetup{justification=raggedright} 	 
	\caption{\label{fig:setup} \textit{Schematic of the experimental chamber setup, showing the two-color XFEL beam incident at normal incidence onto the silicon target. The diffraction detector at the rear side shows example data at 0$\,$fs, with both pulses scattering from the ambient lattice, and at 150$\,$fs, where the delayed probe pulse signal is significantly stronger due to increased diffuse scattering. The image is adapted from our previous paper \cite{Hartley2019}.}}
\end{figure}

The data presented in this paper was collected at BeamLine 3 of the SACLA X-ray Free Electron Laser facility in Hy\={o}go, Japan \cite{Yabashi2015,Tono2013}. Using their two color mode, pulses at different X-ray energies with a variable delay between them are generated by undulators upstream and downstream of an electron chicane \cite{Hara2013}. Both pulses are 7$\,$fs in duration (full width half maximum), such that we can assume that each pulse does not affect the sample structure as it probes it \cite{Inoue2016,Chapman2014}. The delay between the pulse can be varied arbitrarily, with a precision of \textless 1 fs, although delays shorter than the pulse lengths would not give meaningful data in this experimental setup. 

\begin{figure}
	\includegraphics[width=0.48\textwidth]{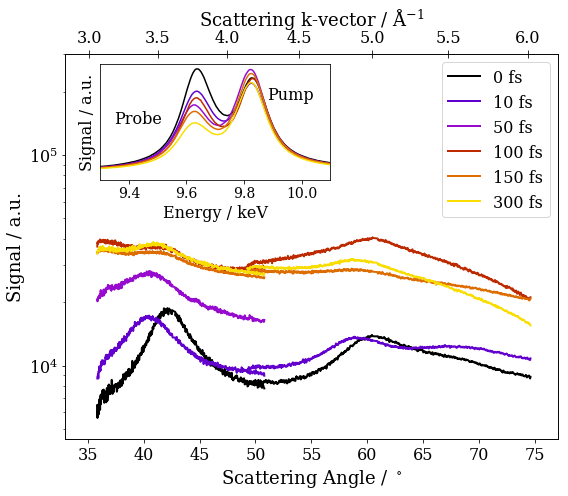} 
	\captionsetup{justification=raggedright} 	 
	\caption{\label{fig:lineouts} \textit{Scattering lineout data from silicon for varying pump-probe delays, aligned such that no Laue spots fall on the detector. XFEL spectra for each delay are shown in the inset. Signals were taken at two separate detector positions, with angular ranges 35\degree-51\degree and 49\degree-75\degree; at 50$\,$fs delay, only the former was taken. Most lineouts are the average of around 300 shots, except 100$\,$fs which uses only 60.}}
\end{figure}

This experiment used an initial (pump) pulse at $9.83\pm0.05\,$keV containing $130\pm40\,\mu$J and a second (probe) pulse at $9.62\pm0.04\,$keV; the probe pulse has a similar fluence at 0 fs delay ($120\pm30\,\mu$J), but this decreases by around 50\% (to $60\pm20\,\mu$J) at the maximum delay of 300$\,$fs, due to loss of electron beam quality in the chicane. A pick-off diffractive optic (\textless1\% loss) upstream of the chamber allows us to monitor the incoming spectrum on each shot, such that it can be accounted for in the analysis \BB{\cite{Tamasaku2015}}; averages of the incoming spectra are shown in the inset of Figure \ref{fig:lineouts}. The beam is focused using Kirkpatrick-Baez mirrors for an elliptical spot size of ($205\pm17\,$nm) $\times$ ($163\pm15\,$nm), which was remeasured by a knife-edge scan after each change in pulse delay. After accounting for losses in the beam propagation and focusing, the spot size gives an incident pump beam intensity (fluence) of 1.6$\pm$0.5$\times10^{19}\,$W/cm$^2$ \allowbreak  (1.1$\pm$0.4$\times10^5\,$J/cm$^2$). Due to the use of reflective, rather than refractive, focusing optics, the spot sizes of the pump and probe beams are almost identical, and there is only minimal ($\sim\!1\%$) signal outside the focal spot \cite{Pikuz2015}.


The pulses were normally incident onto 20 $\mu$m thick, $\langle001\rangle$-oriented single crystalline silicon samples at a repetition rate of 30 Hz, with the target scanned \textgreater 50 $\mu$m between shots to ensure a fresh sample. An estimated 15\% of the pump pulse energy is absorbed by the silicon, exciting photoelectrons which deposit energy across a wide region of the sample \cite{Grum-Grzhimailo2017}, with a maximum range estimated at around $1\,\mu$m \cite{Berger2005}. If we assume that all deposited energy remained within the focal spot, we obtain a deposited dose of around 1.02$\,$keV/atom, while assuming that the energy is uniformly spread across the photoelectron range gives a much lower dose, on the order of 8.3$\,$eV/atom; in practice, the dose in the center of the spot would be somewhat above this lower estimate. There exists significant work estimating the behavior of silicon under XFEL irradiation, with a damage threshold for nonthermal melting estimated at between 0.9$\,$eV/atom \cite{Medvedev2015,Medvedev2019}, with significant interplay between nonthermal and electron-phonon effects, and 2.1$\,$eV/atom \cite{Stampfli1990} under the Born-Oppenheimer approximation. The deposited dose here is well above these estimates, so we expect to see purely nonthermal melting effects, resulting in a high-density liquid within at most 100s of fs, and subsequent ablation over longer timescales \cite{Medvedev2018a}. These rapid phase changes are driven by induced Coulomb forces between the ions, which energetically favor a transition from the inefficiently-packed diamond lattice to a closer-packed liquid \cite{Lyon2015}, since this reduces the average separation, releasing electrostatic potential energy as kinetic energy of the particles.

The scattered signal of both the pump and probe pulses is collected on a pair of MultiPort Charged Couple Devices (MPCCDs) \cite{Kameshima2014}, located outside the chamber at a distance of 169 mm above the XFEL axis. The total detector coverage was $2048 \times 512$ pixels ($102.4 \times 25.6\,$mm$^2$, pixel size $50 \times 50 \,\mu$m$^2$) and could be translated horizontally, along the direction of the beam axis, to obtain data from different angular ranges. The detector positions were calibrated using diffracted signal from copper samples, and covered regions of 35\degree-51\degree and 49\degree-75\degree. Due to the crystal orientation of the sample, no diffraction (Laue) spots can be reached on the detector.

\vspace{5mm}

\begin{figure}
	\includegraphics[width=0.48\textwidth]{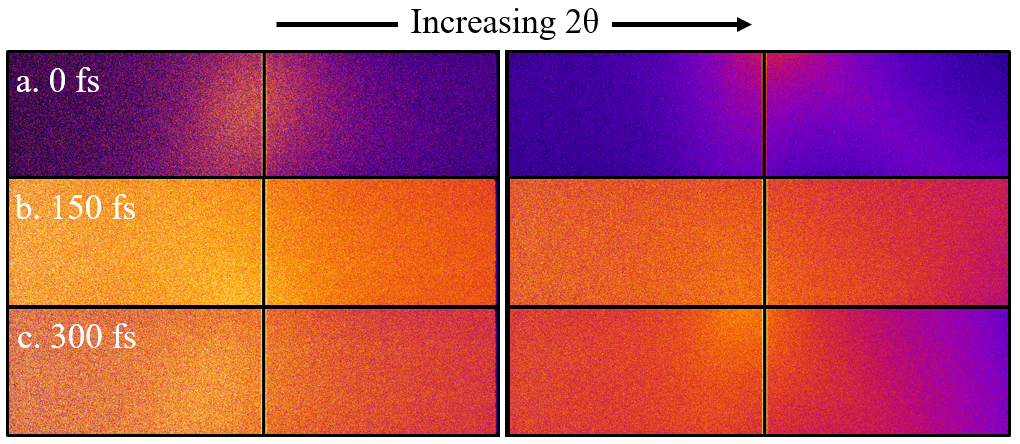} 
	\captionsetup{justification=raggedright} 	 
	\caption{\label{fig:data} \textit{Averaged diffraction data images across the angular ranges, at delays of 0, 150 and 300$\,$fs (top to bottom) between the pump and probe beams. The color scale is arbitrary, but consistent between images.}}
\end{figure}

Example 2D data images shown in Figure \ref{fig:data} are averages over all shots at the same time delay and detector position, and each consist of signal from both the pump and probe beams, which scatter from the ambient and heated sample, respectively.  The pump-probe jitter is \textless 1$\,$fs, although the use of non-gated detectors means that the signal is effectively averaged over the transit time of the pulse through the target, approximately 66$\,$fs. The data images\BB{, and the subsequent lineouts, are adjusted to} account for filtering on the detector and the variable \BB{solid angle} coverage of the pixels.

Because this experimental setup does not probe any coherent diffraction orders, at 0$\,$fs (Figure \ref{fig:data}a) only diffuse signal is seen. This is primarily temperature diffuse scattering (TDS) of both pulses from phonon modes of the crystalline lattice \cite{Warren1969} but may also include signal from low level imperfections and impurities \cite{Barabash2001, Klang2006}. At longer delays, the pump pulse still interacts with the ambient lattice, so a similar (although weaker) TDS signal can be seen in the 300$\,$fs image, on top of a much higher scattered signal of the probe pulse from the heated sample. The significantly weaker TDS at 150$\,$fs appears to be due to a different crystal rotation around the $\langle001\rangle$ vector for this sample. The drop in signal from 150 to 300$\,$fs is primarily due to the weaker probe pulse at the longest delay. 

Figure \ref{fig:lineouts} shows azimuthally integrated lineouts of each delay probed as a function of the scattering angle, along with the associated incoming spectra; the scattering vector (top axis) is calculated for the probe pulse wavelength $\lambda=1.29\,$\AA. The early time lineouts at 0 and 10$\,$fs show weak diffuse scattering, with the difference between them due a slight misalignment of the samples, such that they probe different azimuthal angular ranges. From these to the smoother lineouts at 100+$\,$fs there is a rapid rise in signal strength, with the data point at 50$\,$fs approximately midway between. This rise in signal, across regions of $k$-space with no diffraction peaks, is strong evidence of a loss of order, and a transition to a disordered, liquid-like state within 100$\,$fs of irradiation. The only other source of signal would be emission from the heated target, but this would not be expected to vary with the pulse delay, as a similar electron temperature is reached in all cases and the detectors are not gated. Fourier transforming the signal lineout to observe the corresponding change in the pair distribution function, as is standard in synchrotron experiments e.g. \cite{Kim2005}, is not possible in this case due to the relatively small $k$-range probed. This small $k$-range is determined by the XFEL, which produces significantly lower pulse fluences for X-ray energies above 10 keV, and the chamber geometry, which does not allow us to access a larger angular range.
\vspace{5mm}

\begin{figure}
	\includegraphics[width=0.48\textwidth]{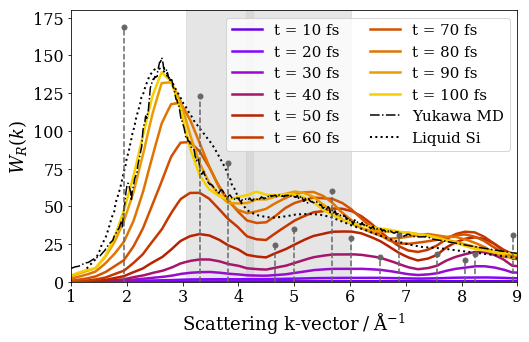} 
	\captionsetup{justification=raggedright} 	 
	\caption{\label{fig:DFTsim} \textit{Simulated Rayleigh weight as a function of scattering vector. The solid lines are extracted from intervals of the DFT-MD simulation, while the dot-dashed line is from a screened Coulomb (Yukawa) MD simulation; all assume a constant ionization of +3 for the form factor. The dashed lines indicate diffraction peaks of the ambient lattice, and are reduced by 75\% for easier comparison. The two shaded areas show the detector angular coverages. The dotted line indicates previously measured scattering from liquid silicon, at an ion temperature of 1893$\,$K \cite{Kimura2001}.}}
\end{figure}

In order to compare the observed phase transition to predicted behavior, we performed a density functional molecular dynamics (DFT-MD) simulation of the crystalline solid with the electron temperature instantaneously raised to $T_e = 10\,$eV, and the sample then allowed to evolve under a microcanonical ensemble. This assumes that there is no bulk change in mass or energy density over the timescale considered, which are justified by the short timescales and the large region heated by the excited photoelectrons, respectively. The atomic positions at each timestep of the simulation were Fourier transformed to obtain the static structure factor $S(k)$ and multiplied by a tabulated atomic form factor $f(k)$, which describes the distribution of electrons around the atom in reciprocal space; in this case, we assumed a constant ionization of +3, estimated from comparison of $S(k)$ with that from a screened Debye model \cite{Vorberger2013}. Together, these give the Rayleigh Weight  $W_R(k) = S(k) \times f(k)^2$, which is directly proportional to the scattered signal.

Figure \ref{fig:DFTsim} shows the simulated Rayleigh Weight as a function of $k$-vector over the delay range 0-100{\,}fs, alongside liquid silicon x-ray diffraction data from a magnetically levitated molten silicon sample at 1893{\,}K \cite{Kimura2001}. Also shown is the output of the simpler molecular dynamics (MD) simulation, which was used to estimate the ionization. The close agreement of this lineout with that of the DFT-MD simulation confirms that the state formed after 100 fs is well described as a one-component plasma interacting through a screened Coulomb potential \cite{Wunsch2008,Vorberger2013}.

The simulated lineouts show a strong rise in signal within 100$\,$fs, in agreement with the observed experimental data. Compared to the liquid signal, the DFT-MD results do not show the `shoulder' at 3.5$\,$\AA$^{-1}$, but have a more pronounced peak at 5.5$\,$\AA$^{-1}$. These differences are likely due to the simulation being ionized, such that the structural behavior is dominated by Coulomb forces rather than transient covalent bonds. This ionization therefore both causes the ultrafast disordering to occur, as described above, but also affects the final state. 

\begin{figure}
	\includegraphics[width=0.48\textwidth]{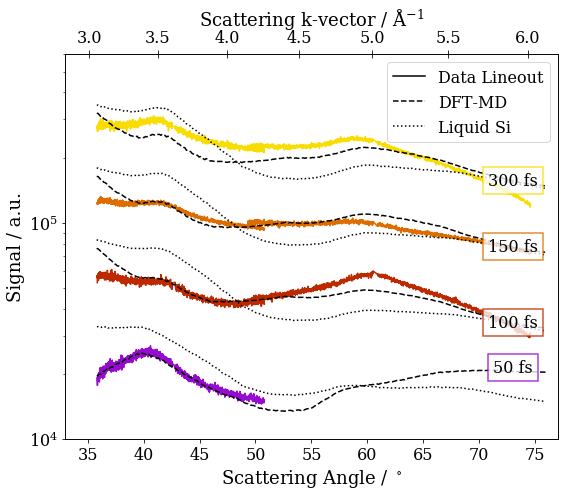} 
	\captionsetup{justification=raggedright} 	 
	\caption{\label{fig:lineouts_sim} \textit{Lineouts from delays of 50, 100, 150 and 300 fs from Figure \ref{fig:lineouts}, compared to the simulated (dashed) and liquid Si (dotted) Rayleigh weights from Figure \ref{fig:DFTsim}. The Rayleigh weights are added to the early time lineout to account for TDS, and scaled by the probe and pump fluence, respectively. The color scale (online only) is the same as in Figure \ref{fig:lineouts}, and the lines are offset for clarity.}}
\end{figure}

 Figure \ref{fig:lineouts_sim} compares the DFT-MD and liquid lineouts to the observed experimental data. The simulated and liquid lines are the sum of the relevant Rayleigh weight and either the $t=0\,$fs or $t=10\,$fs delay signal from Figure \ref{fig:lineouts}, to account for TDS which is not captured in the simulation. The two components are weighted by the relevant probe and pump fluences, respectively, and include an empirical factor to account for the uncalibrated signal units, which is constant for all delays.

From the comparison in Figure \ref{fig:lineouts_sim}, we can clearly see that the simulated DFT-MD data describes the observed disordered structure much better than the liquid silicon data of Kimura \ea\;\cite{Kimura2001}. At all four delays, the liquid underestimates the signal at higher angles, while the Coulomb-dominated liquid of the simulation gives qualitative agreement. The biggest disparity is at the lowest $k$-values, where the simulation predicts a sharp rise in signal for $k\,$\textless$\,3\,$\AA$^{-1}$, which does not appear in the experimental data. Although the uncertainty in the signal is the greatest in this region, since the shielding in the experimental geometry was thickest, this does suggest the simulation may not be fully capturing the behavior of the sample induced by the X-ray heating. Additionally, the weaker agreement at the latest delay implies that further evolution occurs after the initial disordering.


\begin{figure}
	\includegraphics[width=0.48\textwidth]{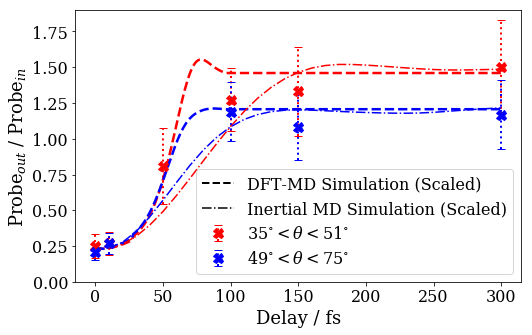} 
	\captionsetup{justification=raggedright} 	 
	\caption{\label{fig:evolution} \textit{Scattered probe signal, normalized to the incoming probe intensity, as a function of time. The simulation lines are scaled to the measured values, and so only demonstrate the evolution timescale.}} 
\end{figure}

To highlight the agreement in the timescale of the disordering, Figure \ref{fig:evolution} plots the scattered probe pulse signal as a function of delay. This is calculated by subtracting the t=0 signal, normalized to the pump pulse intensity of each shot, and normalizing the sum of this pump-subtracted signal to the probe pulse intensity. Also shown is the signal simulated by DFT-MD, described above, and that from a purely inertial MD model -- this simulation was performed with \textsc{Lammps} \cite{LAMMPS,Plimpton1995}, using 1000 Si atoms initialized in a cubic diamond structure at 300$\,$K, and evolved with only a Lennard-Jones potential, such that their evolution is due to room temperature thermal motion \cite{Wang2020}. As expected, the DFT-MD simulation disorders more quickly, due to the induced Coulomb forces driving them towards the disordered state \cite{Lyon2015}. We also note that the timescales for the two angular regions are more similar for the DFT-MD than the inertial case, where the intensity evolves as $I(t) \sim \exp(-k^2 v_{rms}^2 t^2 / 3)$ \cite{Lindenberg2005}, with $v_{rms}=\sqrt{\frac{3 k_B T_0}{m_{Si}}}=5\times10^{-3}\,$\AA/fs the root mean square thermal velocity.

The DFT-MD simulation clearly gives better agreement with the observed behavior, while that predicted by the inertial model lies outside the experimental uncertainty of the 50$\,$fs delay data point. This suggests that the purely inertial model used to explain semiconductor disordering by lower intensity optical pulses \cite{Lindenberg2005} cannot explain that induced by the higher intensity drive here. However, the two effects are only relatively weakly distinguished, and the angular dependence implies that a future experiment looking at lower angles, where inertial behavior would take longer to cause disordering, could further confirm these results.

\vspace{5mm}

We have observed a rapid rise in diffuse scattering from an initially single-crystalline silicon sample irradiated by an intense X-ray pulse, with the resulting disordered state showing a distinctly different structure to that of liquid silicon. This indicates an ultrafast phase change due to Coulomb forces induced between the atoms by ionization. Rather than relying on weakening of diffraction lines, our method observes the rise in diffuse scattering, allowing the structure to be observed throughout and beyond the phase transition, with the potential for unprecedented insight into melting dynamics in a wide range of sample materials. Due to the weak scattering from liquids, this method will generally only be feasible with the high X-ray flux available on XFEL facilities, particularly if applied to even lower Z samples, with carbon likely to beparticularly interesting \cite{Medvedev2018a}. 

In principle, similar methods can be used whenever the scattering from the ambient structure is well-characterized. However, confidence in the measurements of the generated state increases as ambient scattering is reduced, such as by masking diffraction spots, optimizing crystal orientation, or using absorption edge filtering before the detector in order to attenuate the pumping pulse. Alternatively, as new detectors are developed with larger dynamic ranges, experiments which observe both the loss of diffraction peaks and the rise in diffuse scattering simultaneously would give unprecedented insight into phase change dynamics. 

\vspace{5mm}
\begin{acknowledgements}
\small{The XFEL experiments were performed at the BL3 of SACLA with the approval of the Japan Synchrotron Radiation Research Institute (JASRI) (Proposal nos. 2017B8075 and 2018A8056). We would like to thank all staff at SACLA for their technical support on the beamtimes, and Dr. T. White for the use of his \texttt{skcalc.py} software. N.J.H. was supported in part by JSPS KAKENHI Grant no. 16K17846. \R{N.J.H. and E.E.M. acknowledge support from ...??}. N.J.H., K.V., A.K.S. and D.K. were supported by the Helmholtz Association under VH-NG-1141. A.V.R. acknowledges support Australian Government through the Australian Research Council’s Discovery Project DP170100131. N.O. was supported in part by JSPS Japan-Australia Open Partnership Joint Research Project and MEXT Q-LEAP Project.}
\end{acknowledgements}


\end{document}